\documentclass{PoS}

\usepackage{amssymb,amsmath,bm,bbm}
\usepackage{epsf,epsfig}
\usepackage{afterpage}
\usepackage{longtable}
\usepackage{cite}
\usepackage{latexsym, mathrsfs}
\usepackage{graphics}
\usepackage{url}
\usepackage{paralist}
\usepackage{bbold}
\newcommand{\be}{\begin{equation}}
\newcommand{\ee}{\end{equation}}
\newcommand{\bea}{\begin{eqnarray}}
\newcommand{\eea}{\end{eqnarray}}
\newcommand{\bec}{\begin{center}}
\newcommand{\eec}{\end{center}}
\newcommand{\nn}{\nonumber}

\title{Thermalization of a boost-invariant non-Abelian plasma: \\Holographic approach  with boundary sourcing}

\ShortTitle{Thermalization of a boost-invariant non-Abelian plasma}

\author{Loredana Bellantuono\\
   Dipartimento di Fisica, Universit\`a  di Bari,via Orabona 4, I-70126 Bari, Italy\\
INFN, Sezione di Bari, via Orabona 4, I-70126 Bari, Italy\\
       E-mail: \email{loredana.bellantuono@ba.infn.it}}
       
       \author{Pietro Colangelo\\
INFN, Sezione di Bari, via Orabona 4, I-70126 Bari, Italy\\
       E-mail: \email{pietro.colangelo@ba.infn.it}}
       
       \author{\speaker{Fulvia De Fazio}\\
        INFN, Sezione di Bari, \\via Orabona 4, I-70126 Bari, Italy\\
        E-mail: \email{fulvia.defazio@ba.infn.it}}

\author{Floriana Giannuzzi\\
   Dipartimento di Fisica, Universit\`a  di Bari,via Orabona 4, I-70126 Bari, Italy\\
       E-mail: \email{floriana.giannuzzi@ba.infn.it}}

\abstract{In a  holographic approach, the evolution of a 4D  strongly coupled non-Abelian plasma towards equilibrium can be studied investigating a 5D gravitational  dual.  The process  driving the  plasma   out-of-equilibrium
can be described by  boundary sourcing, a deformation of the boundary metric;  the analysis of the late-time dynamics allows to understand how the hydrodynamic regime settles in.  We apply the method to a boost-invariant case, considering the  effects of different quenches, solving
the  Einstein equations in the bulk and studying the time-dependence of observables such as the effective temperature, the energy density and the pressures.  
  The main outcome is that, if the effective temperature of the system when the quench is switched off is   $T_{eff}(\tau^*)=500$ MeV, thermalization is reached within a  time of ${\cal O}$(1 fm/c),
an important information if the case of the QCD plasma produced in relativistic heavy ion collisions is considered. }

\FullConference{The European Physical Society Conference on High Energy Physics\\
		22--29 July 2015\\
		Vienna, Austria}

\begin{document}

\section{Introduction}
For strongly coupled fluids, as those produced in ultrarelativistic heavy ion collisions  at RHIC and  LHC, 
 it is important to understand whether the late-time dynamics  can be described by a low viscosity hydrodynamics, and what is  the elapsed time needed to
 reach  equilibrium after a perturbation. This requires studying  the stress-energy tensor 
$ T^\mu_\nu \sim diag(-\epsilon, p_\perp,p_\perp, p_\parallel)$,
whose components are     the system energy density $\epsilon$ and  the  pressures $p_\perp,  p_\parallel$ along one of the two transverse directions (with respect, e.g., to the heavy ion collision axis)   and in the longitudinal direction.
%(I  refer  to the components of $ T^\mu_\nu$ without considering  the overall factor $\frac{N_c^2}{2 \pi^2}$).
A  framework  to study  the transient process from a far-from-equilibrium state  is the holographic approach, based on the conjectured
correspondence   between  a supersymmetric, conformal field theory defined in a four dimensional (4D) space-time and a  gravity theory in a   five dimensional  Anti de-Sitter space (AdS$_5$) times a  five dimensional sphere  S$^5$ \cite{Maldacena:1997re}.
The  gauge theory is defined on the 4D  boundary of AdS$_5$.  The connection has been extended to the  nonlinear fluid dynamics \cite{Bhattacharyya:2008jc,CasalderreySolana:2011us}.
 Non-equilibrium  can be studied solving   the 5D Einstein bulk equations for the metric subject to suitable time-dependent boundary condition,  and
  the boundary theory stress-energy tensor results  from the holographic renormalization procedure \cite{deHaro:2000xn}.
  
To drive the  system out-of-equilibrium within this framework,   a distortion of the boundary metric due to external sources can be introduced   \cite{Chesler:2008hg}.
 Here we describe the analysis  in  \cite{Bellantuono:2015hxa}, that goes further the proposal in \cite{Chesler:2008hg} in two respects. First,  various kinds of 
 quenches  are considered, to investigate whether and how the onset of the hydrodynamic regime is related to the distortion profiles. Moreover, an algorithm for the numerical solution of the Einstein equations has been developed, the details of which can be found in \cite{Bellantuono:2015hxa}.
 In the holographic  framework, thermalization can also be studied starting from initial states characterized by an assigned energy density \cite{Jankowski:2014lna}.

  \section{Boundary sourcing}
  As described in  \cite{Chesler:2008hg}, the effect of a time-dependent deformation of the 4D boundary metric is the production of  gravitational radiation propagating in the bulk, with the possible formation of a black hole. Hence, the study of the process towards equilibrium in the 4D theory can be afforded studying the time evolution of the 5D dual geometry.
We choose
 4D coordinates  $x^\mu=(x^0,x^1,x^2,x^3)$,   with  $x^3=x_\parallel$ identified with the axis in the direction of collisions and along which  the plasma expands. Boost-invariance along that axis is imposed, as well as   translation invariance and $O(2)$ rotational invariance  in the     $x_\perp=\{x^1,\,x^2 \}$ plane. In terms of  the proper time $\tau$ and rapidity $y$,   with $x^0=\tau \cosh y$, $x_\parallel=\tau \sinh y$,  the 4D Minkowski line element  reads:
$ds^2_4=-d\tau^2+dx_\perp^2+\tau^2 dy^2$. 

A way of distorting the boundary metric,  leaving the spatial three-volume unvaried and respecting the  chosen symmetries, is through a function   $\gamma(\tau)$:
\be
ds^2_4=-d\tau^2+e^{\gamma(\tau)} dx_\perp^2+\tau^2e^{-2\gamma(\tau)} dy^2 \,\,. \label{metric4D}
\ee
(\ref{metric4D}) can be viewed as the metric of  the boundary of a 5D bulk in which the gravity dual  is defined. Adopting  Eddington-Finkelstein  coordinates  and denoting with $r$ the fifth coordinate, the 5D bulk metric can be written  as
\be
ds^2_5=2 dr d\tau-A d\tau^2+ \Sigma^2 e^B dx_\perp^2+ \Sigma^2 e^{-2B}dy^2  \,\,\, . \label{metric5D}
\ee
The boundary corresponds to $r \to \infty$.
Due to the imposed symmetries, the metric functions
 $A$, $\Sigma$ and $B$   depend  on $r$ and $\tau$ only.  Directional derivatives along the infalling radial null geodesics and the outgoing radial null geodesics can be defined,  $\xi^\prime=\partial_r \xi$ and ${\dot \xi}= \partial_\tau \xi+\frac{1}{2} A \partial_r \xi
$, for  a generic  function $\xi(r,\tau)$.

The 5D metric (\ref{metric5D}) solves    Einstein equations with negative cosmological constant. In terms of $A(r,\tau)$, $\Sigma(r, \tau)$ and  $B(r,\tau)$   the equations can be cast in the form   \cite{Chesler:2008hg}:
\bea
&&\Sigma ({\dot \Sigma})^\prime +2 \Sigma^\prime {\dot \Sigma}-2 \Sigma^2=0  \label{ein1} \\
&& \Sigma ({\dot B})^\prime+\frac{3}{2} \left(\Sigma^\prime {\dot B}+B^\prime {\dot \Sigma}\right)=0  \label{ein2}\\
&& A^{\prime \prime} +3 B^\prime {\dot B} -12 \frac{\Sigma^\prime {\dot \Sigma} }{\Sigma^2}+4=0\label{ein3} \\
&& {\ddot \Sigma}+\frac{1}{2} \left( {\dot B} ^2 \Sigma -A^\prime  {\dot \Sigma} \right) =0\label{ein4} \\
&& \Sigma^{\prime \prime}+\frac{1}{2} B^{\prime 2} \Sigma =0 \label{ein5}\,\,. \eea
The solutions are  constrained to reproduce
 the 4D metric  (\ref{metric4D}) for $r \to \infty$, hence
 %This constrains  the  $r\to \infty$ behavior of  $A(r,\tau)$, $\Sigma(r, \tau)$ and  $B(r,\tau)$:
\be
\frac{A(r,\tau)-r^2 }{r}\xrightarrow[r\to \infty] { } 0 \,\,, \hskip 1 cm 
 \frac{\Sigma(r,\tau)}{r} \xrightarrow[r\to \infty]  { } \tau^{\frac{1}{3}}\,\,, \hskip 1 cm  
B(r,\tau)\xrightarrow[r\to \infty] { }-\frac{2}{3} \log \tau +\gamma(\tau)  \,\, .\label{large-r1} 
\ee
Moreover,  if the 
 distortion of the boundary metric starts at $\tau=\tau_i$,  
the metric functions at $\tau_i$ must  give  
 the  AdS$_5$ metric
\be
ds^2_5= 2 dr d\tau + r^2 \left[ -d \tau^2+ d x_\perp^2 + \left( \tau+\frac{1}{r} \right)^2 dy^2 \right]   \,\,\, , \label{AdS5}
\ee
hence   initial conditions must be imposed,
\be
A(r,\tau_i) = r^2  \,\,, \hskip 1 cm
\Sigma (r,\tau_i)=r \left(\tau_i+\frac{1}{r} \right)^{1/3}  \,\,, \hskip 1 cm
B(r,\tau_i)=-\frac{2}{3} \log \left( \tau_i+\frac{1}{r} \right) .
\ee
Various profiles can be chosen for the deformation $\gamma(\tau)$, with different  structures, duration and intensity.
We generically define the  profile
\be
\gamma(\tau) =w \left[ Tanh \left(\frac{\tau -\tau_0}{\eta} \right) \right]^7 \,+\sum_{j=1}^{N}\gamma_j(\tau,\tau_{0,j})  \,\,\, , \label{profile}
\ee
with
\be
\gamma_j(\tau,\tau_{0,j}) = c_j f_j(\tau,\tau_{0,j})^6 e^{-1/f_j(\tau,\tau_{0,j})} \Theta\left(1-\frac{(\tau-\tau_{0,j})^2}{\Delta_j^2}\right) ,\hskip 0.3cm
f_j(\tau,\tau_{0,j})= 1- \frac{(\tau-\tau_{0,j})^2}{\Delta_j^2} \,\,\, .\label{profile2}
\ee
 This represents sequences of $N$ "pulses",  each one of intensity $c_j$ and duration proportional to $\Delta_j$, with the possibility of superimposing a smooth deformation of intensity $w$. The response of the $5D$ bulk geometry to the quenches describes the way  equilibration is achieved.

In \cite{Bellantuono:2015hxa}  different models have been investigated,   corresponding to different sets of   parameters in Eqs.~(\ref{profile})-(\ref{profile2}).
We report the results of model  $\cal A$ , representing two pulses,  with parameters 
$w=0$ and $N=2$, 
$c_1 = 1$, $\Delta_{1,2}=1$, $\tau_{0,1}=\frac{5 }{4}\Delta_1$,
$c_2 = 2$. Moreover,   two different values of $\tau_{0,2}$:   $\tau_{0,2}=\frac{17}{4} \Delta_2$ and $\tau_{0,2}=\frac{9}{4} \Delta_2$ are considered, varying the time elapsed between the pulses.  The  distortion ends at   $\tau_f^{\cal A}=5.25$ and  $\tau_f^{\cal A}=3.25$ for the two cases, respectively.
Model $\cal C$  is a sequence of pulses superimposed to a smooth distortion, and is obtained with
\bea
w&=& \frac{2}{5}, \,\,\,\,\, \eta=1.2 , \,\,\,\,\, \tau_0=0.25 \,\,\,\,\,\,
\Delta_j=0.8 \,\,\,\,\, ( j=1,2,3,4)\,\,, 
c_{0,1}= 0.5, \,\,\,\,\,
c_{0,2}= 1, \,\,\,\,\,
c_{0,3}= 0.35, \,\,\,\,\, \nn \\
c_{0,4}&=& 0.3,  \,\,\,\,\,
\tau_{0,1}=  \tau_0+1.5, \,\,\,\,\,
\tau_{0,2}=  \tau_0+3.8, \,\,\,\,\,
\tau_{0,3}=  \tau_0+6.1, \,\,\,\,\,
\tau_{0,4}=  \tau_0+8.4\,\,. \nn 
\eea
In this case  the last short pulse ends at $\tau_f^{\cal C}=9.45$.

Black-brane solutions of the Einstein equations for the various  models are found numerically \cite{Bellantuono:2015hxa}, obtaining 
  the apparent  and the event horizon allowing to  define the system effective temperature  $T_{eff}$, together with
$\Sigma(r_h,\tau)^3$,  the area of each horizon per unit of rapidity. The components of the boundary stress-energy tensor are also obtained.

Assuming homogeneity,  boost-invariance and invariance under rotations  in the transverse plane with respect to the fluid velocity, the components of  $T_{\mu}^{\nu}$   only depend on the proper time $\tau$ \cite{Bjorken:1982qr}. Moreover,  $T_{\mu}^{\nu}$ conserved and  traceless implies that  a single function $f(\tau)$ is involved, so that one writes 
\be
T_{\mu}^{\nu}=diag \left(-f(\tau),\,f(\tau)+\displaystyle{\frac{1}{2}}\tau f^\prime(\tau),\,f(\tau)+\displaystyle{\frac{1}{2}}\tau f^\prime(\tau),\, -f(\tau)-\tau f^\prime(\tau)\right) \,\,\,\ .
\ee
For a  perfect fluid,  the equation of state $\epsilon=3 p$,  with $p=p_\parallel=p_\perp$,   fixes the $\tau$ dependence: $\epsilon(\tau)=\displaystyle{\frac{const}{\tau^{4/3}}}$, while
viscous effects produce a deviation  of  $f(\tau)$
 \cite{Janik:2005zt}.
An effective temperature $T_{eff}(\tau)$ can be defined through the  relation
$
\epsilon(\tau)= \displaystyle{\frac{3}{4}} \pi^4 T_{eff}(\tau)^4 $.
The calculation in ${\cal N}=4$  SYM gives the asymptotic $\tau$ dependence \cite{Heller:2007qt}:
%\cite{Heller:2007qt,Baier:2007ix,Heller:2012je}:
\bea
T_{eff}(\tau)&=&\frac{\Lambda}{(\Lambda \tau)^{1/3}} \Bigg[ 1-\frac{1}{6 \pi (\Lambda \tau)^{2/3}}+\frac{-1+\log 2}{36 \pi^2 (\Lambda \tau)^{4/3} } \nn \\
&+&\frac{-21+2\pi^2+51 \log 2 -24 (\log 2)^2}{1944 \pi^3 (\Lambda \tau)^2} + {\cal O}\left( \frac{1}{(\Lambda \tau)^{8/3}} \right )\Bigg] \,\,\ , \label{Teff1}
\eea
corresponding, for the energy density,   the pressure ratio and anisotropy, to
\bea
\epsilon(\tau)&=& \frac{3 \pi^4 \Lambda^4}{4 (\Lambda \tau)^{4/3} }\left[ 1-\frac{2c_1}{ (\Lambda \tau)^{2/3}}+\frac{c_2}{ (\Lambda \tau)^{4/3}} + {\cal O}\left( \frac{1}{(\Lambda \tau)^2} \right )\right] \label{hydroeps} \\
\frac{p_\parallel}{p_\perp}&=& 1-\frac{6c_1}{ (\Lambda \tau)^{2/3}}+\frac{6c_2}{ (\Lambda \tau)^{4/3}} + {\cal O}\left( \frac{1}{(\Lambda \tau)^2} \right) ,\label{pressure-ratio} \\
\frac{\Delta p}{\epsilon} &=& \frac{p_\perp-p_\parallel}{\epsilon} =2\left[ \frac{c_1}{ (\Lambda \tau)^{2/3}}+\frac{2c_1^2-c_2}{ (\Lambda \tau)^{4/3}} +{\cal O}\left( \frac{1}{(\Lambda \tau)^2} \right) \right] \,\,\, \label{pressure-anis}
\eea
($c_1=\displaystyle{\frac{1}{3 \pi}}$ and $c_2=\displaystyle{\frac{1+2 \log{2}}{18 \pi^2}}$). The parameter $\Lambda$  depends on the  model for quenches.

\section{Results}
Although  we discuss here   the results  of models ${\cal A}$ and   ${\cal C}$,
all the considered models share common features, namely the formation of a horizon and the evolution towards the hydrodynamic regime \cite{Bellantuono:2015hxa}. Such evolution has different features for the various observables.
The effective temperature and the energy density start evolving according to their hydrodynamic expression immediately after the quench is switched off. On other hand, pressures take longer to reach the hydrodynamic regime:  a "thermalization time" $\tau_p$ can be defined from the condition  
\be
\Big|\frac{p_{||}(\tau_p)/p_\perp(\tau_p)-(p_{||}(\tau_p)/p_\perp(\tau_p))_H}{p_{||}(\tau_p)/p_\perp(\tau_p)}\Big| =0.05 \,\,\, , \label{taup}
\ee
 where $(p_{||}(\tau_p)/p_\perp(\tau_p))_H$ is given by (\ref{pressure-ratio}).
\begin{figure}[t!]
\bec
\begin{tabular}{lll}\hspace*{-0.5cm}
\includegraphics[width = 0.35\textwidth]{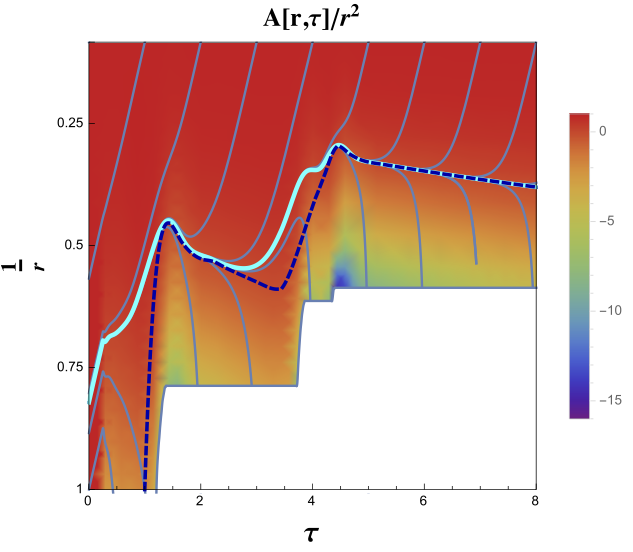}&\hspace*{-0.7cm}
\includegraphics[width = 0.35\textwidth]{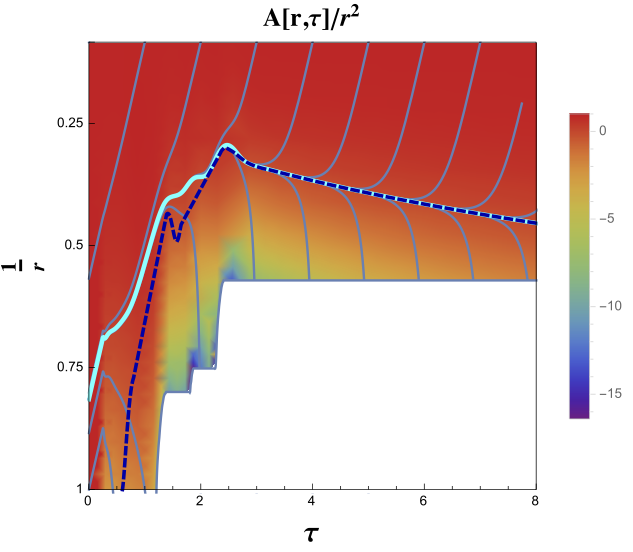}&\hspace*{-0.7cm}
\includegraphics[width = 0.35\textwidth]{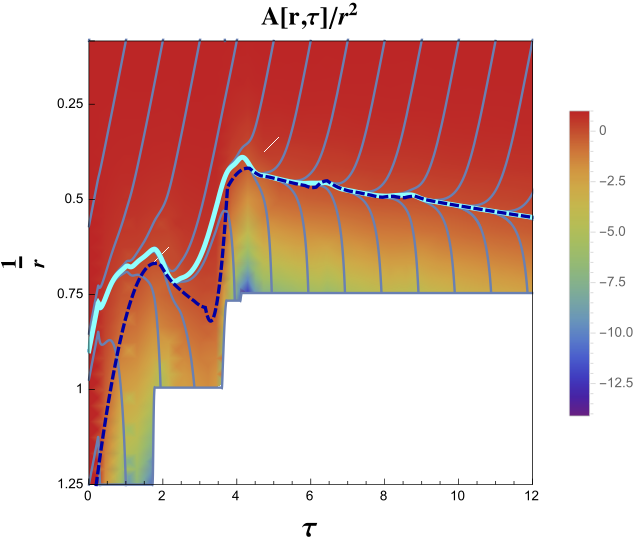}
\end{tabular}
\caption{\small  The  panels show the function $A(r,\tau)/r^2$ vs $\tau$ and $1/r$,   obtained solving  the Einstein equations  in the case of the model $\cal A$ 
 of two distant (left) or close pulses (central panel)  in the boundary metric, and of the model $\cal C$ (right panel). The color bars indicate the values of the function. The gray lines are radial null outgoing geodesics,  the dashed  dark blue line is the apparent  horizon,   the continuous cyan line is  the event horizon. The  excision in the low-$r$ region used in the calculation  is  shown.}
\label{density-due-picchi}
\eec
\end{figure}

\begin{figure}[t!]
\bec
\begin{tabular}{lll}
\includegraphics[width = 0.35\textwidth]{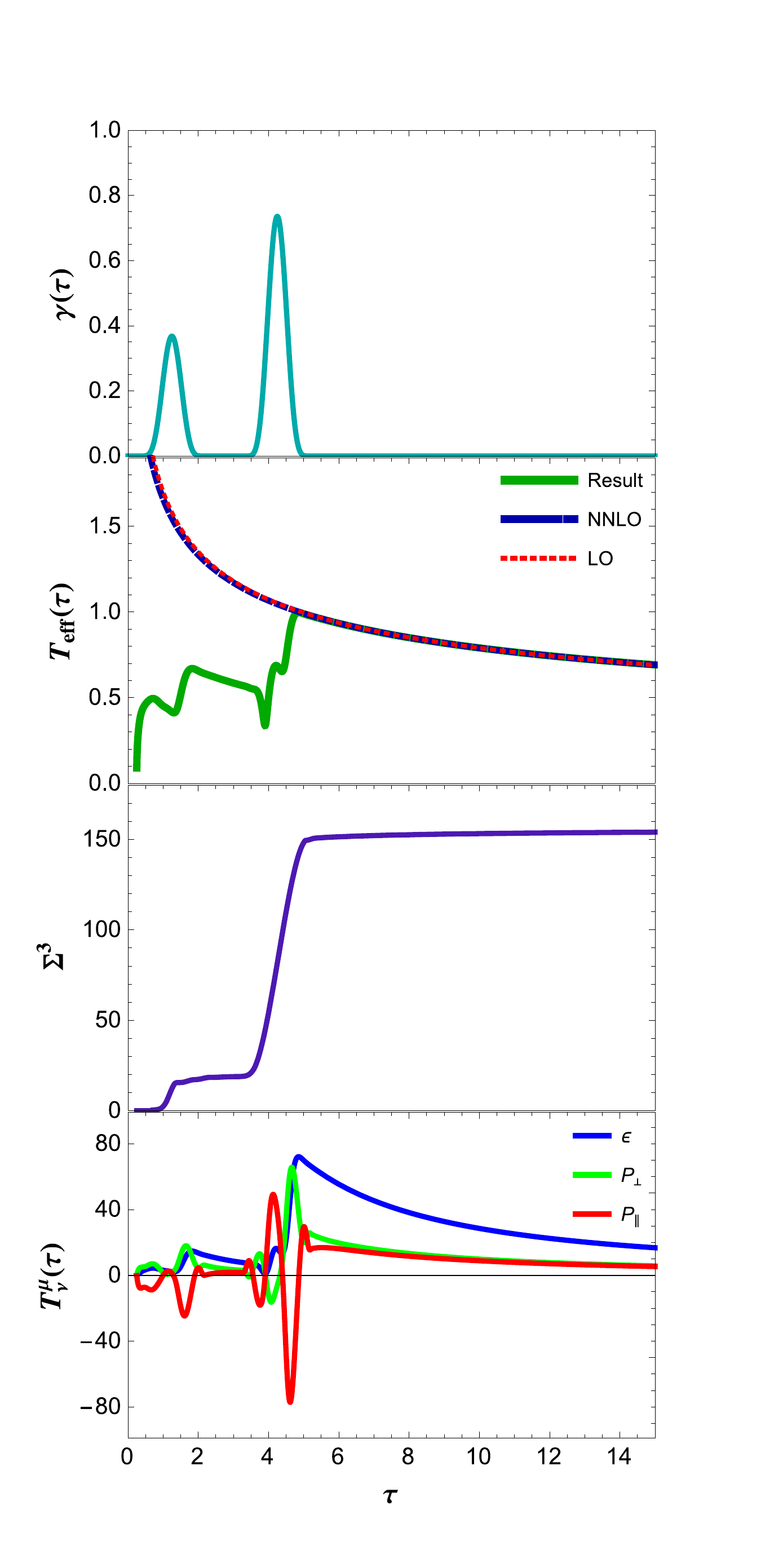}&\hspace*{-1.cm}
\includegraphics[width = 0.35\textwidth]{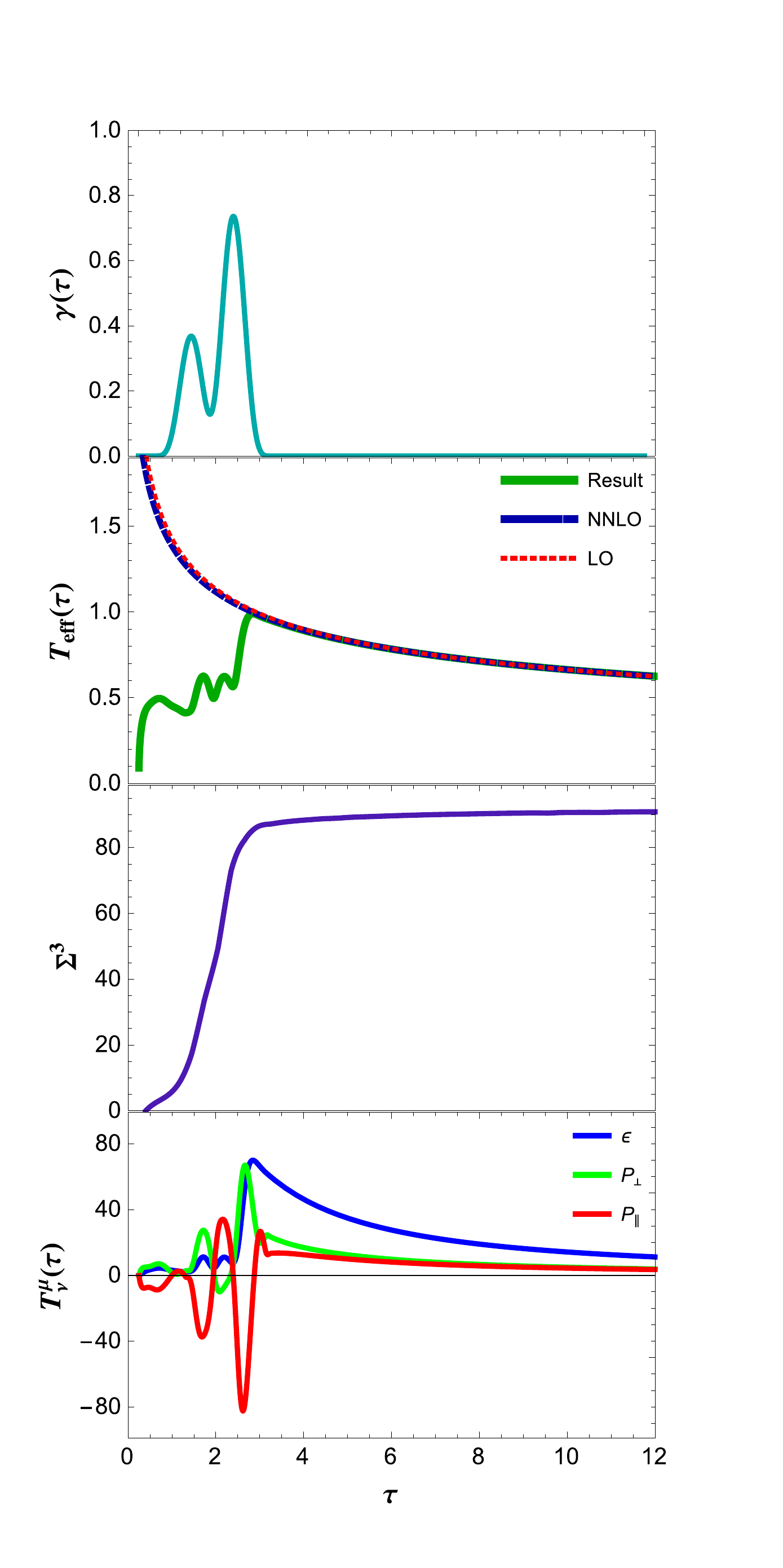} &\hspace*{-1.cm}
\includegraphics[width = 0.35\textwidth]{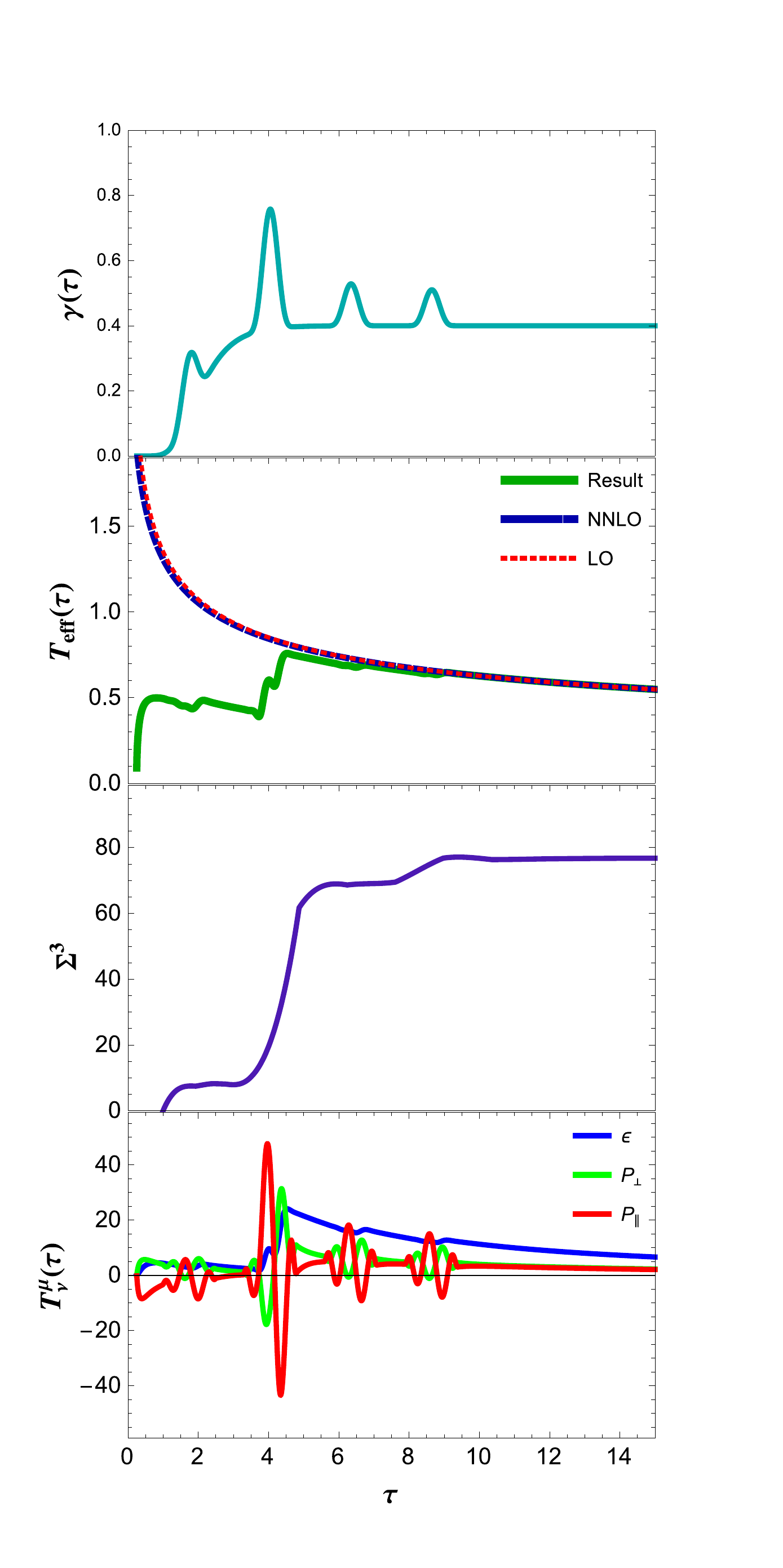} 
\end{tabular}
\vspace*{-0.5cm}
\caption{\small  For three  different quench profiles (models $\cal A$  and $\cal C$) the panels show (from top to bottom)  the profile $\gamma(\tau)$, the   temperature $T_{eff}(\tau)$,  the horizon area $\Sigma^3(r_h,\tau)$ per unit of rapidity,  and the three components  $\epsilon(\tau)$, $p_\perp(\tau)$ and $p_\parallel(\tau)$ of the stress-energy tensor $T^\mu_\nu$.}
\label{Column2p}
\eec
\end{figure}

In the case of  model ${\cal A}$,  for two distant pulses   the horizon forms in a two-step process in which  the sequence of the boundary distortion is followed, as shown in Fig.~\ref{density-due-picchi} (left panel) which displays the metric  function
$A(r,\tau)/r^2$, together with the apparent and  event horizons. The temperature, reported in  Fig.~\ref{Column2p}, has two decreasing regimes,
each one after each of the two  pulses, with the   relaxation of the system starting  immediately after the end of each perturbation. 
The same can be noticed looking at   the area of the apparent horizon per unit of rapidity, $\Sigma(r_h,\tau)^3$,   which  reaches a constant value, soon after the pulses. On such time scales, the behaviour of the components of $T_{\nu}^{\mu}$ shows  no isotropization  after the first pulse.

For  two  nearly overlapping quenches, no  structure is observed before the end of  the perturbation. The horizon area per unit of rapidity  sharply increases during the distortion, reaching the asymptotic large $\tau$ value $\Sigma(r_h,\tau)^3= \pi^3 \Lambda^2$ soon after the end of the distortion.

The time $\tau^*$ at which energy density  $\epsilon(\tau^*)$ differs from the hydrodynamic value $\epsilon_H(\tau^*)$  in  (\ref{hydroeps})
by less than  $5 \%$ essentially coincides with the end of the perturbation, while the time defined in (\ref{taup}) is larger.
The  difference $\tau_p-\tau^*$ measures  the elapsed time between  the end of the quench and the restoration of  the hydrodynamical regime. It can be expressed in physical units introducing  a scale in the system by imposing 
$T_{eff}(\tau^*)=500$ MeV. 
The  time scales and the  values of the parameter $\Lambda$   are collected in Table
\ref{tab}. The difference $\tau_p-\tau^*$ is  found to be order (or less than)  1 fm/c, an important information for the strongly coupled QCD plasma. This result favourably compares with QCD-based models \cite{Ruggieri:2015yea}.
\vspace*{0.5cm}
\begin{figure}[thb!]
\bec
\begin{tabular}{lll}
\includegraphics[width = 0.35\textwidth]{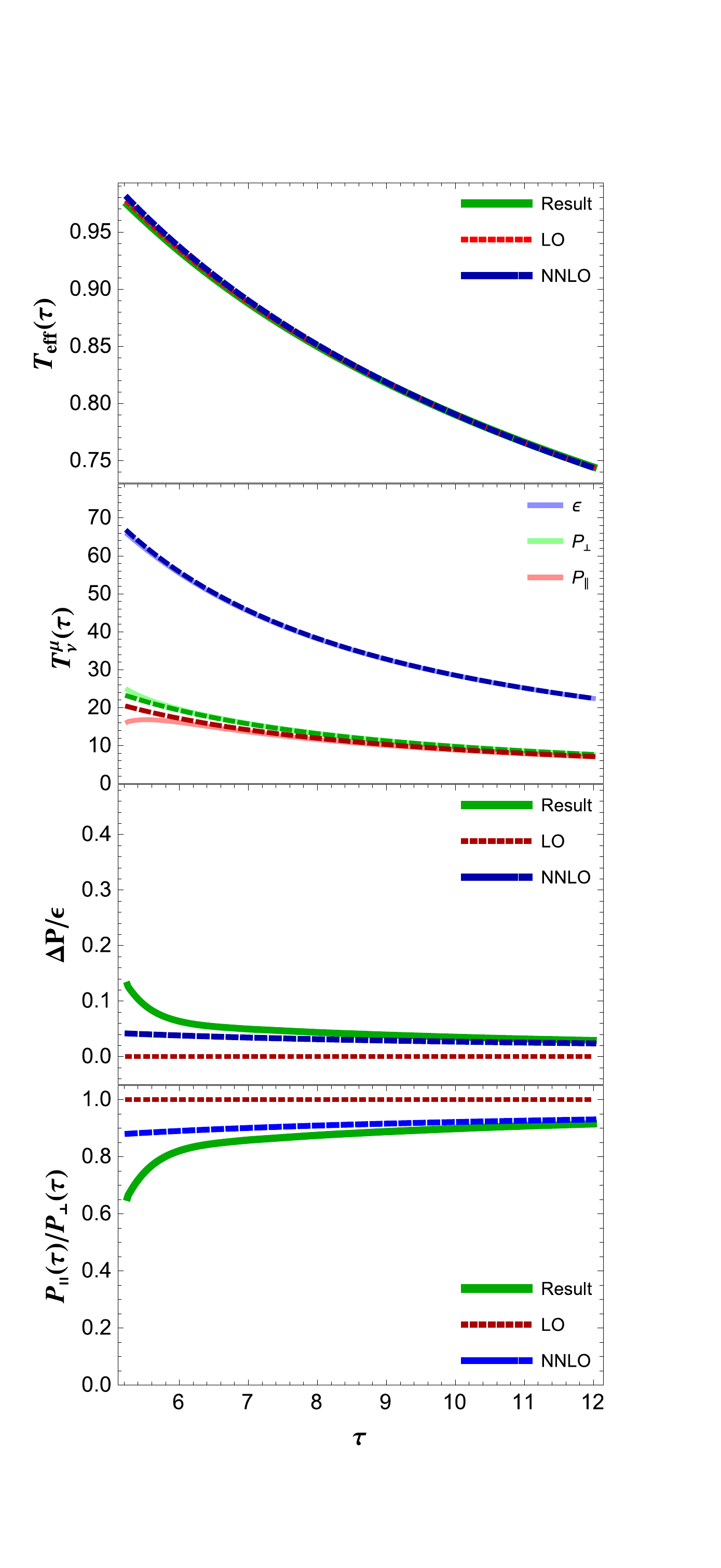}&\hspace*{-1.cm}
\includegraphics[width = 0.35\textwidth]{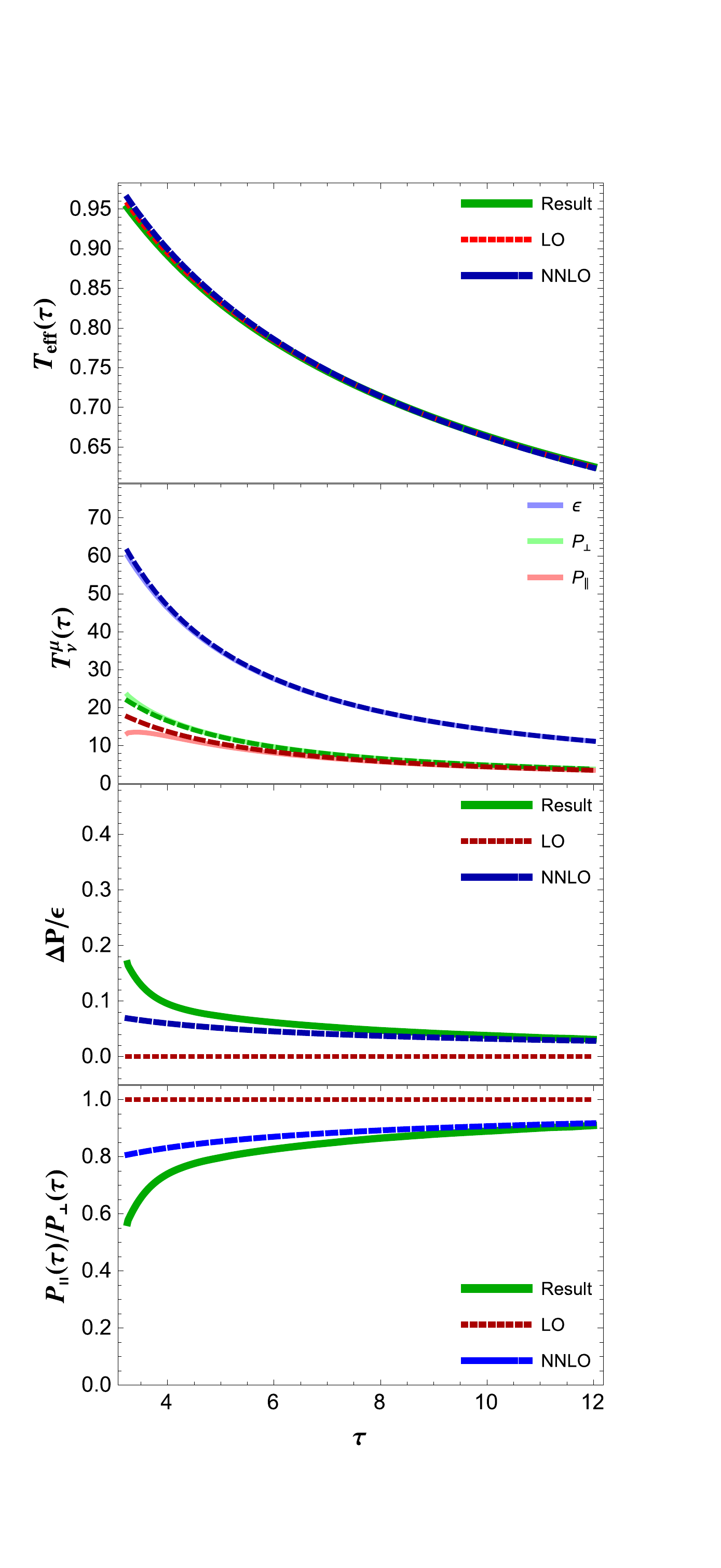} &\hspace*{-1.cm} 
\includegraphics[width = 0.35\textwidth]{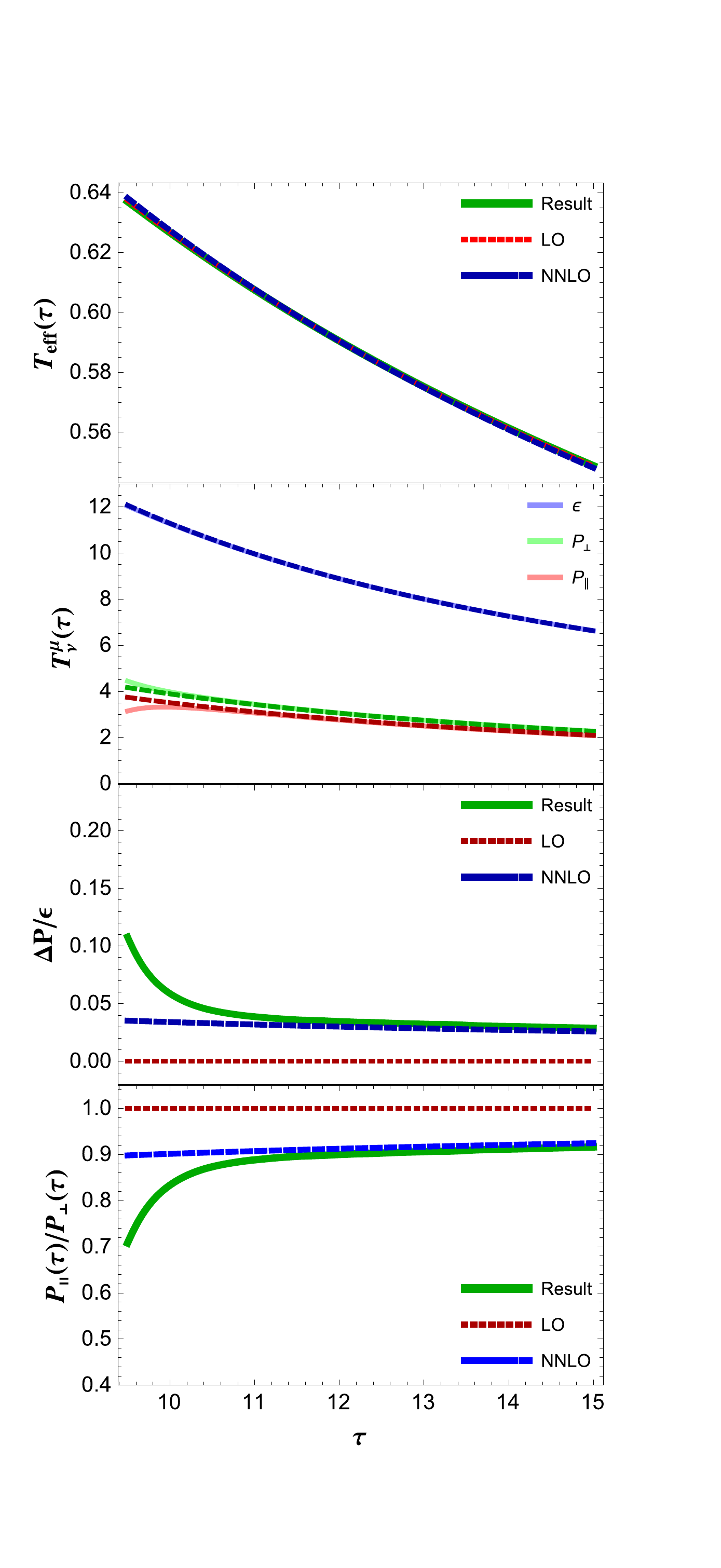}
\end{tabular}
\eec
\vspace*{-1.5cm}
\caption{\small   For three  different quench profiles (models $\cal A$  and $\cal C$) the panels show,  from top to bottom: temperature $T_{eff}(\tau)$,    components  $\epsilon(\tau)$, $p_\perp(\tau)$ and $p_\parallel(\tau)$ of the stress-energy tensor,
pressure anisotropy  $\Delta p/\epsilon=(p_\perp-p_\parallel)/\epsilon$ and ratio $p_\parallel/p_\perp$, computed for $\tau > \tau_f^{\cal A, \cal C}$.  The short and long dashed lines correspond to the hydrodynamic result and to the NNLO result in the $1/\tau$ expansion. }\label{Column2p-a}
\end{figure}
\begin{table}[h]
\bec
\begin{tabular}{| c c c c c |}
\hline
model& $\tau^*$ & $\tau_p$ & $\Lambda $ & $\Delta \tau= \tau_p-\tau^*$ (fm/c)\\ \hline 
 ${\cal A}~(1)$ & 5.25 & 6.8 & 2.25 & 0.60 \\ 
 ${\cal A}~(2)$ & 3.25 & 6.0 & 1.73 & 1.03 \\ 
% ${\cal B}$ & 5 & 6.74 & 1.12& 0.42 \\ 
 ${\cal C}$ & 9.45 & 10.24 & 1.59 & 0.20 \\ \hline 
  \end{tabular}
\eec
\caption{Results for different  models of boundary quenches.}\label{tab}
\end{table}

\section{Conclusions}
Through  distortions of the boundary metric, in a boost-invariant setup and using holography, it is possible to describe relaxation of an out-of-equilibrium strongly coupled fluid. 
 Independently of the  considered deformation, we have found that
a horizon forms  in the bulk metric, and an effective
$\tau-$dependent temperature can be defined. 
 $T_{eff}(\tau)$  evolves according to a viscous hydrodynamic expression  as soon as the quench  is switched off, while the pressures have an elapsed time
  (setting $T_{eff}(\tau^*)=500$ MeV) of a fraction of fm/c  before the hydrodynamic regime is reached.

\end{document}